\documentclass[twocolumn,showpacs,preprintnumbers,amsmath,amssymb,nofootinbib]{revtex4}
\input psfig.sty

\usepackage{graphicx}
\usepackage{dcolumn}
\usepackage{bm}

\begin{document}

\title{Interpreting Cosmological Vacuum Decay}

\author{J. S. Alcaniz{\footnote{alcaniz@on.br}}}

\affiliation{Departamento de Astronomia, Observat\'orio Nacional,
20921-400 Rio de Janeiro - RJ, Brasil}

\author{J. A. S. Lima{\footnote{limajas@astro.iag.usp.br}}}

\affiliation{Instituto de Astronomia, Geof\'{\i}sica e Ci\^encias
Atmosf\'ericas, USP, 05508-900 S\~ao Paulo, SP, Brasil}

\date{\today}

\begin{abstract}

The cosmological vacuum decay scenario recently proposed by Wang
and Meng \cite{wm} is rediscussed. From thermodynamic arguments it
is found that the $\epsilon$ parameter quantifying the vacuum
decay rate must be positive in the presence of particle creation.
If there is no particle creation, the proper mass of Cold Dark
Matter (CDM) particles is necessarily a time dependent quantity,
scaling as $m(t) = m_o a(t)^{\epsilon}$. By considering the
presence of baryons in the cosmological scenario, it is also shown
that their dynamic effect is to alter the transition redshift
$z_*$ (the redshift at which the Universe switches from
decelerating to accelerating expansion), predicting values of
$z_*$ compatible with current estimates based on type Ia
supernova. In order to constrain the $\Omega_m - \epsilon$ plane,
a joint statistical analysis involving the current supernovae
observations, gas mass fraction measurements in galaxy clusters
and CMB data is performed. At 95\% c.l. it is found that the
vacuum decay rate parameter lies on the interval $\epsilon = 0.11 \pm 0.12$). The possibility of a vacuum decay into
photons is also analyzed. In this case, the energy density of the
radiation fluid scales as $\rho_r = \rho_{ro}a^{-4 + \epsilon}$,
and its temperature evolution law obeys $T(t) =
T_oa(t)^{\epsilon/4 - 1}$.

\end{abstract}

\maketitle

\section{Introduction}

There is nowadays significant observational evidence that the
expansion of the Universe is undergoing a late time acceleration
\cite{perl,wmap,rnew,allen,revde}. This, in other words, amounts
to saying that in the context of Einstein's general theory of
relativity some sort of \emph{dark energy}, constant or that
varies only slowly with time and space, dominates the current
composition of the cosmos (see, e.g., \cite{revde} for some recent
reviews on this topic). The origin and  nature of such an
\emph{accelerating field} constitutes a completely open question
and represents one of the major challenges not only to cosmology
but also to our current understanding of fundamental physics.

Among many possible alternatives, the simplest and most
theoretically appealing possibility for dark energy is the energy
density stored on the true vacuum state of all existing fields in
the Universe, i.e., $\rho_{\Lambda} = \Lambda/8 \pi G$, where
$\Lambda$ is the cosmological constant. From the observational
side, flat models with a relic cosmological term  ($\Lambda$CDM)
seems to be in agreement with almost all cosmological
observations, which makes them an excellent description of the
observed universe. From the theoretical viewpoint, however, the
well-known cosmological constant problem, i.e., the unsettled
situation in the particle physics/cosmology interface, in which
the cosmological upper bound ($\rho_{\Lambda} \lesssim 10^{-47}
\rm{GeV}^4$) differs from theoretical expectations
($\rho_{\Lambda} \sim 10^{71} \rm{GeV}^4$) by more than 100 orders
of magnitude, originates an extreme fine-tuning problem
\cite{revde1} or makes a complete cancellation (from an unknown
physical mechanism) seem more plausible.

In this regard, a phenomenological attempt at alleviating such a
problem is allowing $\Lambda$ to vary\footnote{Strictly speaking,
in the context of classical general relativity any additional
$\Lambda$-type term that varies in space or time should be thought
of as a new \emph{time-varying field} and not as a cosmologial
constant. Here, however,  we adopt the usual nomenclature of
time-varying or dynamical $\Lambda$ models.}. Cosmological
scenarios with a time-varying or dynamical $\Lambda$ were
independently proposed almost twenty years ago in Refs.
\cite{ozer} (see also \cite{bron}). Afterward, a number of models
with different decay laws for the variation of the cosmological
term were investigated in Ref. \cite{lambdat0} and the
confrontation of their predictions with observational data has
also been analyzed by many authors \cite{lambdat1}. It is worth
mentioning that the most usual critique to these $\Lambda$(t)CDM
scenarios is that in order to establish a model and study their
observational and theoretical predictions, one needs first to
specify a phenomenological time-dependence for $\Lambda$. In this
concern, an interesting step towards a more realistic decay law
was given recently by Wang \& Meng in Ref. \cite{wm}. Instead of
the traditional approach, they deduced a new decay law from a
simple argument about the effect of the vacuum decay on the cold
dark matter (CDM) expansion rate. Such a decay law is similar to
the one originally obtained in Ref. \cite{shapiro1} from arguments based on
renormalization group and seems to be very general, having many of
the previous attempts as a particular case and being capable of
reconciling $\Lambda$(t)CDM models with an initially decelerated
and late time accelerating universe, as indicated by current SNe
Ia observations \cite{rnew}.

The aim of the present paper is is twofold: first, to interpret
thermodynamically the process of cosmological vacuum decay, as
suggested in Ref. \cite{wm}. From thermodynamic considerations, it
is shown that such a process leads to two different effects,
namely, a continuous creation of particles and an increasing in
the mass of CDM particles given by $m(t) = m_o a(t)^{\epsilon}$,
where $a(t)$ is the cosmological scale factor and $\epsilon$ is
the parameter quantifying the decay vacuum rate; second, to
analyze the dynamic modifications in the original Wang-Meng cosmic
scenario by introducing explicitly the baryonic component. As we
shall see, the presence of baryons alters considerably the
accelerating redshift $z_*$, that is, the redshift at which the
Universe switches from deceleration to acceleration. In order to
constrain the parametric space $\Omega_m - \epsilon$, we also
perform a statistical analysis involving three sets of
observables, namely, the latest Chandra measurements of the X-ray
gas mass fraction in 26 galaxy clusters, as provided by Allen et
al. \cite{allen}, the so-called ``gold" set of 157 SNe Ia,
recently published by Riess et al. \cite{rnew}, and the
measurement of the CMB shift parameter, as given by WMAP, CBI, and
ACBAR \cite{wmap}. Finally,  we extend the treatment of Ref.
\cite{wm} to a scenario in which the vacuum energy decays into
photons. In this case, it is found that the temperature evolution
law of radiation is modified to $T = T_oa(t)^{\epsilon/4 - 1}$.

\section{Vacuum Decay into CDM}

Let us first consider the Einstein field equations
\begin{equation}
R^{\mu \nu} - \frac{1}{2}Rg^{\mu \nu} = \chi \left[T^{\mu \nu} +
\frac{\Lambda} {\chi}g^{\mu \nu}\right],
\end{equation}
where $R^{\mu \nu}$ and $R$ are, respectively, the Ricci tensor
and the scalar curvature, $T^{\mu \nu}$ is the energy-momentum
tensor of matter fields and CDM particles, and $\chi =  8 \pi G$
($c = 1$) is the Einstein's constant. Note that according to the
Bianchi identities, the above equations implies that $\Lambda$ is
necessarily a constant either if $T^{\mu \nu} = 0$ or if $T^{\mu
\nu}$ is separately conserved, i.e., $u_{\mu}T^{\mu \nu};_{\nu} =
0$. In other words, this amounts to saying that (i) vacuum decay
is possible only from a previous existence of some sort of
non-vanishing matter and/or radiation, and (ii) the presence of a
time-varying cosmological term results in a coupling between
$T^{\mu \nu}$ and $\Lambda$. For the moment, we will assume a
coupling only between vacuum and CDM particles, so that
\begin{equation}
\label{coupling}  u_{\mu}{\cal{T}}^{\mu \nu};_{\nu} =
-u_{\mu}(\frac{\Lambda g^{\mu \nu}}{\chi});_{\nu},
\end{equation}
or, equivalently,
\begin{equation}\label{coupling1}
\dot \rho_m + 3\frac{\dot a}{a}\rho_m \ = - \dot \rho_v,
\end{equation}
where $\rho_m$ and $\rho_v$ are the energy densities of the CDM
and vacuum, respectively, and  ${\cal{T}}^{\mu \nu} = \rho_m
u^{\mu} u^{\nu}$ denotes the energy-momentum tensor of the CDM
matter.

As commented earlier, the traditional approach for $\Lambda$(t)CDM
models was first to specify a phenomenological decay law and then
establish a cosmological scenario (see, e.g.,
\cite{lambdat0,lambdat1}). Here, however, we follow the arguments
presented in Ref. \cite{wm}, in which a decay law is deduced from
the effect it has on the CDM evolution. The qualitative argument
is the following: since vacuum is decaying into CDM particles, CDM
will dilute more slowly compared to its standard evolution,
$\rho_m \propto a^{-3}$. Thus, if the deviation from the standard
evolution is characterized by a positive constant $\epsilon$,
i.e.,
\begin{equation}
\label{energyCDM} \rho_m=\rho_{mo} a^{-3 + \epsilon},
\end{equation}
Eq. (\ref{coupling1}) yields
\begin{equation}\label{decayv}
\rho_{v} =  \tilde\rho_{vo} + \frac{\epsilon \rho_{m0}}{3 -
\epsilon}a^{-3 + \epsilon},
\end{equation}
where $\rho_{mo}$ is the current CDM energy density and $
\tilde\rho_{vo}$ stands for what is named in Ref. \cite{wm} ``the
ground state value of the vacuum". As discussed there, such a
decay law seems to be the most general one, having many of the
previous phenomenological attempts as a particular case.

\section{Thermodynamics of vacuum decay}

Let us now investigate some thermodynamic features of the decaying
vacuum scenario described in the last section.  As discussed in
Ref. \cite{Lima1}, the thermodynamic behavior of a decaying vacuum
system is simplified if one assumes that the chemical potential of
the vacuum component is zero, and also if the vacuum medium plays
the role of a condensate carrying no entropy, as happens in the
two fluid description employed in superfluid thermodynamics. In
this case, the thermodynamic description require only the
knowledge of the particle flux, $N^{\alpha} = nu^{\alpha}$, and
the entropy flux, $S^{\alpha}=n\sigma u^{\alpha}$, where $n =
N/a^{3}$ and $\sigma = S/N$ are, respectively, the concentration
and the specific entropy (per particle) of the created component.

It is clear from last Section that in the Wang-Meng description
the two component are changing energy, but it is not clear where
the vacuum energy is going to or, in other words, where the CDM
component is storing the energy received from the vacuum decay
process. In principle, since the energy density of the cold dark
matter is $\rho = nm$, there are two possibilities:

\vspace{0.2cm}

{\bf{(i)}} the equation describing concentration, $n$, has a
source term while the proper mass of CDM particles remains
constant;

\vspace{0.3cm}

{\bf{(ii)}} the mass $m$ of the CDM particles is itself a
time-dependent quantity while the total number of CDM particles,
$N = na^{3}$, remains constant.

\vspace{0.2cm}

The case (i) seems to be physically more realistic, and coincides
exactly with the description presented in Ref. \cite{Lima1}.
However, for the sake of completeness, in what follows we consider
both cases.

\subsection{Case I: Vacuum decay into CDM particles}

In this case, there is necessarily a source term in the current of
CDM particles, that is,  $N^{\alpha};_{\alpha} = \psi$. In terms
of the concentration it can be written as
\begin{equation}\label{conc}
\dot n + 3\frac{\dot a}{a}n \ = \psi = n\Gamma,
\end{equation}
where $\psi$ is the particle source ($\psi > 0$), or a sink ($\psi
< 0$), and we have written it in terms of a decay rate, $\Gamma$.
Since $\rho = nm$ we find from (\ref{energyCDM}) that $n =
n_{o}a^{-3 + \epsilon}$.  Inserting this result into the above
equation it follows that

\begin{equation}\label{gamma}
\Gamma = \epsilon \frac{\dot a}{a}.
\end{equation}

The vacuum decay and the associated particle creation rate are the
unique sources of irreversibility. Thermodynamically, the overall
energy transfer from  the vacuum to the fluid component may
happens in several ways. In the most physically relevant case it
has been termed adiabatic decaying vacuum \cite{Lima1}(see also
\cite{pavon} for more applications of adiabatic decay processes in
cosmology). In this case, several equilibrium relations are
preserved, and, perhaps, more important, the entropy of the
created particles increases but the specific entropy (per
particle) remains constant ($\dot \sigma = 0$). This means that

\begin{equation}
\frac{\dot S}{S} = \frac{\dot N}{N} = \Gamma.
\end{equation}
On the other hand, from Eq. (\ref{gamma})
we see that the total number of particles scales as a power law
\begin{equation}\label{TN}
N(t) = N_o a(t)^{\epsilon},
\end{equation}
whereas the second law of thermodynamics, {$\dot S \geq 0$},
implies that  $\epsilon \geq 0$, as should be expected. To close
the connection with the Wang-Meng scenario we need to show that
the vacuum energy density follows naturally from the thermodynamic
approach. Actually, for an adiabatic vacuum decay process  one may
write (see Eqs. (8) and (19) of Ref. \cite{Lima1})

\begin{figure*}
\vspace{0.2in}
\centerline{\psfig{figure=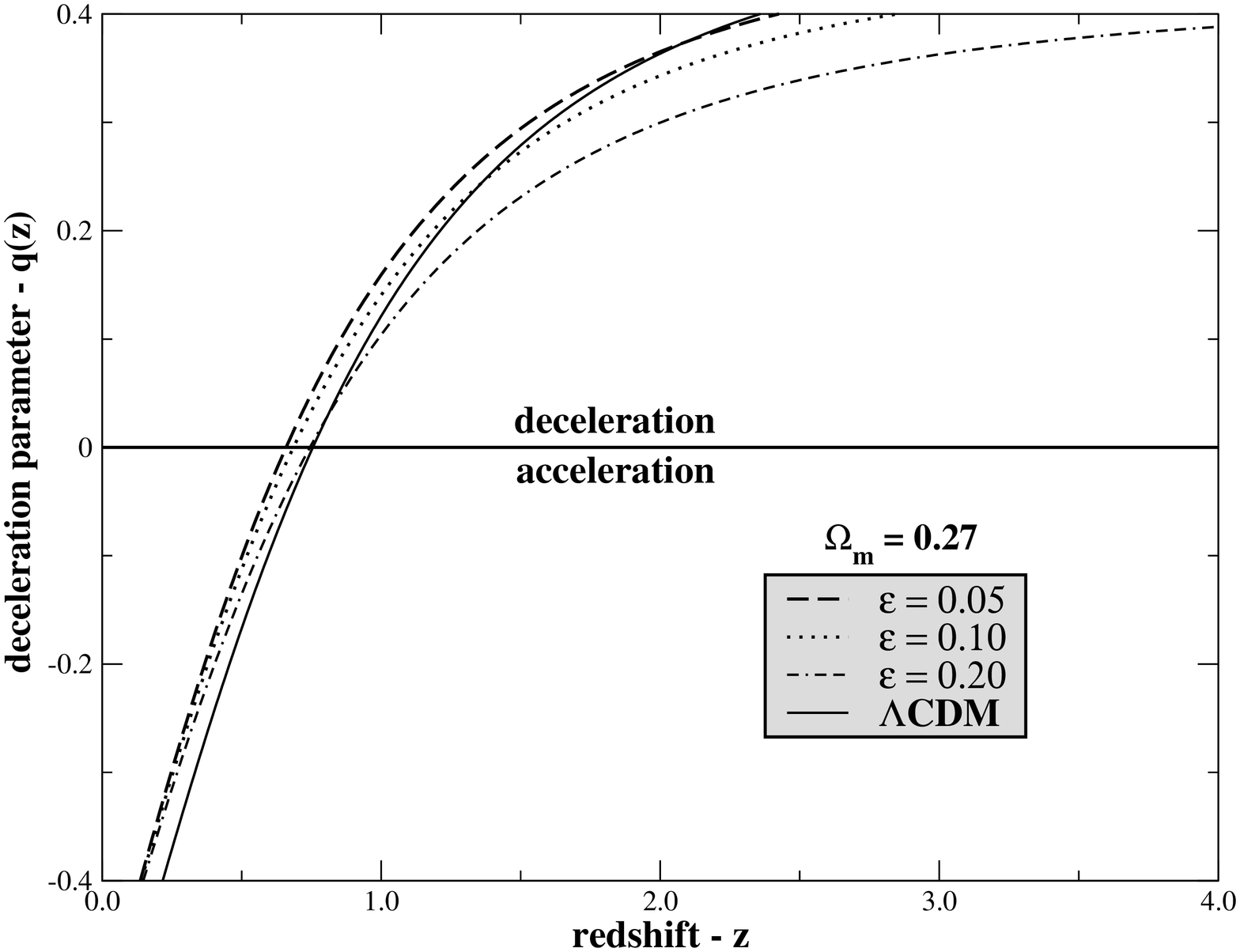,width=2.4truein,height=2.7truein,angle=-90}
\psfig{figure=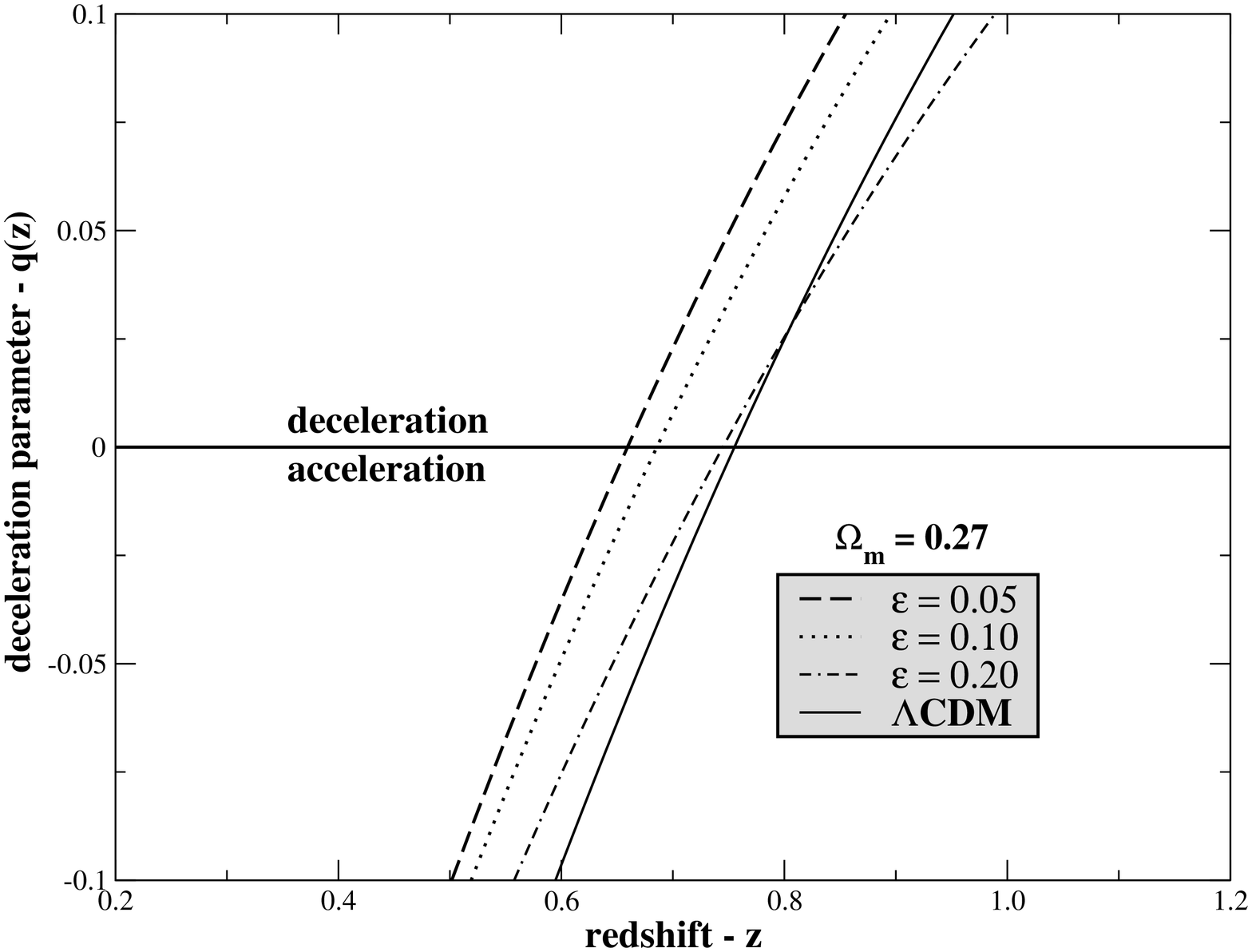,width=2.4truein,height=2.7truein,angle=-90}
\psfig{figure=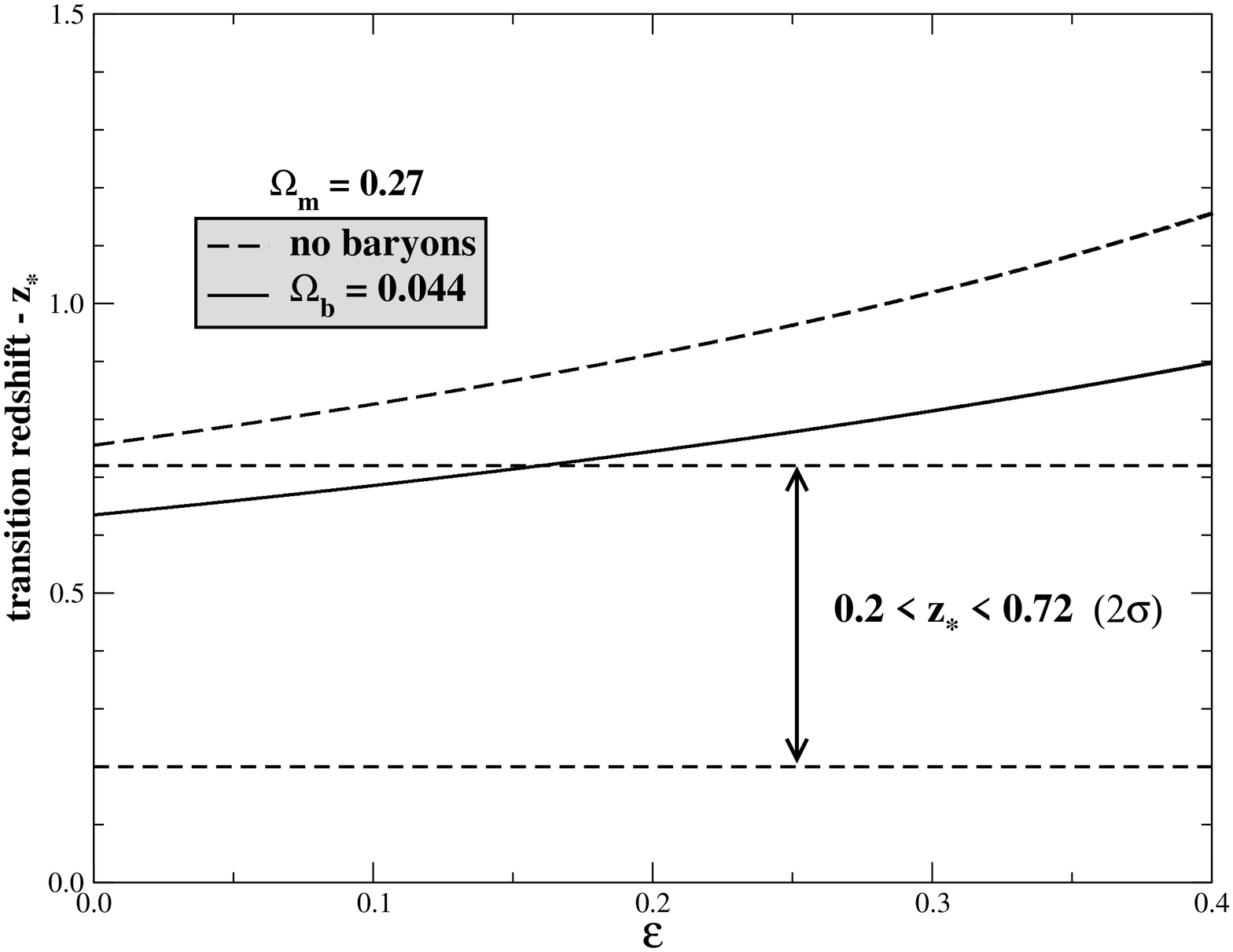,width=2.4truein,height=2.7truein,angle=-90}
\hskip 0.1in} \caption{Effect of baryons on the transition epoch.
{\bf{a)}} The deceleration parameter as a function of redshift for
some selected values of $\epsilon$. In all curves a baryonic
content corresponding to $\simeq 4.4\%$ of the critical density
has been considered. {\bf{b)}} A closer look at Panel (a).
{\bf{c)}} The transition redshift $z_*$ as a function of the decay
rate parameter $\epsilon$. The two cases displayed correspond to
the scenario discussed in Ref. \cite{wm} (``no baryons") and the
scenario proposed in this paper ($\Omega_b = 0.044 \pm 0.004$).
The horizontal dashed lines stand for the 2$\sigma$ interval $0.2
\leq z_* \leq 0.72$, as provided by SNe Ia observations
\cite{rnew}. Note that the unique way to make vacuum decay models
compatible with the SNe Ia interval for $z_*$ is to consider
explicitly the presence of baryons (see Eq. \ref{friedmann}). In
particular, from this analysis we find $\epsilon \leq 0.16$.}
\end{figure*}

\begin{equation}\label{AD}
\dot \rho_v = - \beta\psi,
\end{equation}
where the phenomenological parameter $\beta$ is defined by
\begin{equation}
\beta = \frac{\rho + p}{n}.
\end{equation}
Finally, by considering that the CDM medium is pressureless, Eq.
(\ref{AD}) can be rewritten as

\begin{equation} \label{rhov1}
\dot \rho_v = - nm \epsilon\frac{\dot{a}}{a},
\end{equation}

or still,

\begin{equation} \label{rhovf}
\dot \rho_v = - \rho_{mo}\epsilon a^{-4 + \epsilon} \dot{a},
\end{equation}
whose integration reproduces expression (\ref{decayv}) previously
derived by Wang and Meng \cite{wm}. Beyond the independent
derivation of the decaying vacuum energy density, the interesting
point here is that the sign of the ``coupling constant",
$\epsilon$, is constrained by the second law of thermodynamics.

\subsection {Case II: Variable Mass Particles}

In this case, \emph{there is no creation of CDM particles}, which
means that the concentration satisfies the equation
\begin{equation}
\label{ndot} \dot n + 3\frac{\dot a}{a}n = 0,
\end{equation}
whose solution is $n=n_{o}a^{-3}$ which implies that $N(t) =
constant$. Naturally, if CDM particles are not being created, the
unique possibility is an increasing in the proper mass of CDM
particles. Actually, since $\rho = nm$, Eqs. (\ref{energyCDM}) and
(\ref{ndot}) imply that the mass of the CDM particles scales as
\begin{equation}
m(t) = m_o a(t)^{\epsilon},
\end{equation}
where $m_o$ is the present day mass of CDM particles (compare with
expression (\ref{TN})). Note that this approach for the vacuum
decay process leads to a VAMP\footnote{VAriable Mass
Particles}-type scenario, in which the interaction of CDM
particles with the dark energy field imply directly in an
increasing of the mass of CDM particles (see, e.g., \cite{vamp}
and references therein for more about VAMP models). To complete
our thermodynamic approach for the vacuum decay, a similar
treatment for the case in which the vacuum decays only into
photons is briefly presented in Appendix A.

\section{Observational Aspects}

In this Section we study some observational aspects of the
cosmological scenario discussed above. The Friedmann equation for
this modified $\Lambda$(t)CDM cosmology reads
\begin{equation}
\label{friedmann} (\frac{{H}}{H_o})^2 = \left[ \Omega_b a^{-3} +
\frac{3\Omega_m}{3 - \epsilon}a^{-3 + \epsilon} +
\tilde{\Omega}_{vo}\right],
\end{equation}
where $\Omega_b$ and $\Omega_m$ are, respectively, the baryon and
CDM density parameters and $\tilde{\Omega}_{vo}$ is the density
parameter associated with ``the ground state of vacuum". Note that
unlike Eq. (6) of Ref. \cite{wm}, the above Friedmann equation has
an additional term which accounts for the baryon contribution to
the cosmic expansion. The presence of such a term -- redshifting
as $(1 + z)^3$ -- is justified here since the vacuum is assumed to
decay only into CDM particles.

\subsection{Transition epoch}

Although subdominant at the present stage of cosmic evolution, the
baryonic content may be important for reconciling $\Lambda$(t)CDM
models with some current cosmological observations. As an example,
let us consider the transition redshift, $z_{*}$, at which the
Universe switches from deceleration to acceleration or,
equivalently, the redshift at which the deceleration parameter
vanishes. From Eq. (\ref{friedmann}), it is straightforward to
show that the deceleration parameter, defined as $q =
-a\ddot{a}/\dot{a}^2$, now takes the following form
\begin{equation} \label{10}
q(a)= \frac{3}{2}\frac{\Omega_ba^{-3} + \Omega_m a^{-3 +
\epsilon}}{\Omega_b a^{-3} + \frac{3\Omega_m}{3 - \epsilon}a^{-3 +
\epsilon} + \tilde{\Omega}_{vo}} - 1,
\end{equation}
where we have set $a_o = 1$.

Two important aspects concerning the above equation should be
emphasized at this point. First, note that the presence in Eq.
(\ref{10}) of a non-null density parameter associated with the
ground state of vacuum makes possible a transition
deceleration/acceleration, as indicated by current SNe Ia
observations \cite{rnew}. As well discussed in Ref. \cite{wm}, in
most of the cases, $\Lambda$(t)CDM models without such a term
predict a universe which is either always accelerating or always
decelerating from the onset of matter domination up to today.
Second, note also that, due to the presence of the baryons, the
transition epoch is delayed relative to previous cases (including
the standard $\Lambda$CDM model), which seems to be in better
agreement with recent results indicating $z_{*} = 0.46 \pm 0.13$
at 1$\sigma$ \cite{rnew}.

To better visualize the effect of baryons on the transition epoch,
we show in Fig. 1a the behavior of the deceleration parameter as a
function of redshift [Eq. (\ref{10})] for selected values of the
parameter $\epsilon$. In agreement with WMAP estimates \cite{wmap}
we also assume $\Omega_m = 0.27 \pm 0.04$ and $\Omega_b = 0.044
\pm 0.004$. The best fit $\Lambda$CDM case (the so-called
"concordance model") is also showed for the sake of comparison.
Note that at late times ($z = 0$), since $\epsilon$ is a positive
quantity, the standard $\Lambda$CDM scenario always accelerates
faster than $\Lambda$(t)CDM models, with the condition for current
acceleration being ${\tilde{\Omega}}_{vo} > \frac{\Omega_b}{2} +
\frac{3\Omega_m(1 - \epsilon)}{6 - 2\epsilon}$. A closer look at
the results shown in Fig. 1a is displayed in Fig. 1b. In Fig. 1c
we show the transition redshift $z_*$ as a function of the
parameter $\epsilon$, which is obtained from the expression
\begin{equation}
\Omega_b (1 + z_*)^3 + \left[\frac{3 - 3\epsilon}{3 -
\epsilon}\right]\Omega_m(1 + z_*)^{3 - \epsilon} -
2{\tilde{\Omega}}_{vo} = 0.
\end{equation}
Two different cases are shown. The scenario of Ref. \cite{wm} (no
baryons -- dashed line) and the model presented here (solid line),
in which the baryonic content accounts for $\sim 4.4\%$ of the
critical density. As physically expected (due to the  attractive
gravity associated with the baryonic content), $z_*$ is always
smaller in the latter scenario than in the former. In particular,
by considering the 2$\sigma$ interval $0.2 \lesssim z_* \lesssim
0.72$ \cite{rnew} (horizontal dashed lines) we find $\epsilon
\lesssim 0.16$, which is in fully agreement with the results of
the statistical analysis performed in the next Section.

\subsection{SNe Ia, Clusters and CMB Constraints}

In order to delimit the parametric space $\Omega_m - \epsilon$ we
perform in this Section a joint statistical analysis involving
three complemetary sets of observations. We use to this end the
latest Chandra measurements of the X-ray gas mass fraction in 26
galaxy clusters, as provided by Allen et al. \cite{allen} along
with the so-called ``gold" set of 157 SNe Ia, recently published
by Riess et al. \cite{rnew}, and the estimate of the CMB shift
parameter \cite{wmap}, $R \equiv
\Omega_m^{1/2}\Gamma(z_{\rm{CMB}}) = 1.716 \pm 0.062$ from WMAP,
CBI, and ACBAR \cite{wmap}, where $\Gamma(z)$ is the dimensionless
comoving distance and $z_{\rm{CMB}} = 1089$. In our analysis, we
also include the most recent determinations of the baryon density
parameter, as given by the WMAP team \cite{wmap}, i.e.,
$\Omega_bh^2 = 0.0224 \pm 0.0009$ and the latest measurements of
the Hubble parameter, $h = 0.72 \pm 0.08$, as provided by the HST
key project \cite{hst} (we refer the reader to \cite{refer} for
more details on the statistical analysis).

In Fig. 2 we show the results of our statistical analysis.
Confidence regions ($68.3\%$, 95.4$\%$ and 99.7$\%$) in the plane
$\Omega_m - \epsilon$ are shown for the particular combination of
observational data described above. Note that, although the limits
on the parameter $\epsilon$ are very restrictive, the analysis
clearly shows that the model presented here constitutes a small
but significant deviation from the standard $\Lambda$CDM dynamics.
The best-fit parameters for this analysis are $\Omega_m = 0.27$
and $\epsilon = 0.11$, with the relative $\chi^2_{min}/\nu \simeq
1.12$ ($\nu$ is defined as degrees of freedom). Note that this
value of $\chi^2_{min}/\nu$ is similar to the one found for the
so-called ``concordance model" by using SNe Ia data only, i.e.,
$\chi^2_{min}/\nu \simeq 1.13$ \cite{rnew}. At 95.4$\%$ c.l. we
also found $\Omega_m = 0.26 \pm 0.05$ and $\epsilon = 0.11 \pm
0.12$.

\begin{figure}
\vspace{.2in}
\centerline{\psfig{figure=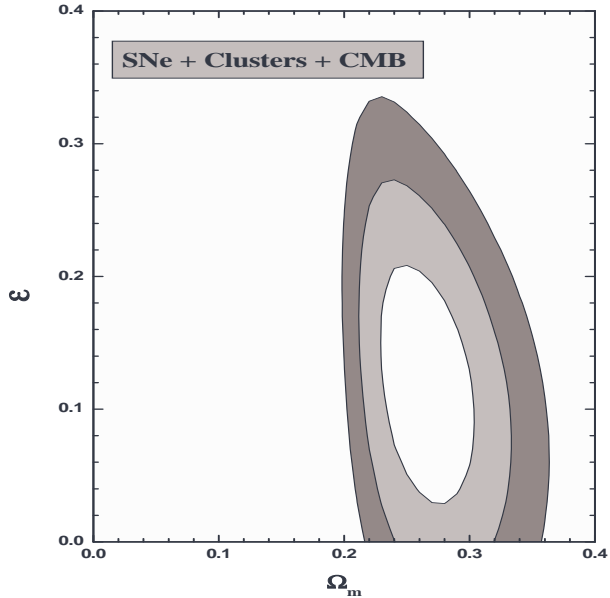,width=3.4truein,height=3.4truein,angle=0}
\hskip 0.1in} \caption{The plane $\Omega_m - \epsilon$ for the
$\Lambda(t)$CDM scenario. The curves correspond to confidence
regions of $68.3\%$, 95.4$\%$ and 99.7$\%$ for a joint analysis
involving SNe Ia, Clusters and CMB data. The best-fit parameters
for this analysis are $\Omega_m = 0.27$ and $\epsilon = 0.11$,
with reduced $\chi^2_{min}/\nu \simeq 1.12$.}
\end{figure}

\section{Conclusion}

In this paper we have slightly modified and interpreted several
features of the decaying vacuum scenario recently proposed by Wang
and Meng \cite{wm}. A baryonic component has been explicitly
introduced, and we have seen that it has an important dynamic
effect, namely, the transition epoch from a
decelerating/acelerating regime is delayed relative to the one
predicted by the original Wang-Meng scenario (including the
standard $\Lambda$CDM model). The importance of the baryonic
contribution cannot be neglected because it reconciles the
decaying vacuum scenario with the recent observations \cite{rnew}
(see figure 1, panel c). However, other details of the radiation
and matter dominated phases are not modified. This is easily
verified by computing the value of the redshift $z$ for which
$\rho_{b} = \rho_m$. For the present values of the density
parameters, $\Omega_{mo}\sim 0.3$ and $\Omega_{bo}=0.04$, one
finds $z \simeq 10^{1/\epsilon}$. Therefore, for $\epsilon \simeq
0.11$ (the best-fit found in this paper), we obtain $z \simeq
10^{10}$. In other words, after this redshift, the Universe is
still radiation dominated but the baryons are already subdominant
in comparison to the CDM component.

We have also discussed some thermodynamic aspects of such a
scenario assuming that the baryonic component is identically
conserved. In particular, if CDM particles are produced by the
decaying vacuum, we shown that the sign of the coupling parameter,
$\epsilon$,  is restricted by the second law of thermodynamics to
assume only positive values. In this case, the total number of CDM
particles is a time-dependent function given by $N(t)
=N_{o}a^{\epsilon}$. However, VAMP-type scenarios - VAriable mass
particles - are also possible when the total number of particles
remains constant. In this case, the mass scales as $m(t) = m_o
a^{\epsilon}$, that is, the energy of the vacuum decay process is
totally transformed in mass of the the existing particles.
Naturally, if photons are produced, the temperature law of
radiation must also be affected. This case has been discussed with
some detail in the Appendix A.

\appendix

\section{Vacuum decay into radiation}

In this Appendix we briefly discuss how the Wang-Meng treatment
can be extended to the case of radiation. Now, the energy
conservation law reads

\begin{equation}
\label{ECL} \dot \rho_r + 4H\rho_r= -\dot \rho_v,
\end{equation}
where $\rho_r$ is the radiation energy density. By considering
that radiation will dilute more slowly compared to its standard
evolution, $\rho_m \propto a^{-4}$, and that such a deviation is
characterized by a positive constant $\alpha$ we find
\begin{equation}\label{energyR}
\rho_r = \rho_{ro} a(t)^{-4 + \alpha},
\end{equation}
where $\rho_{ro}$ is the present day energy density of radiation.
For an adiabatic vacuum decay the equilibrium relations are
preserved \cite{Lima1,pavon}, as happens with the Stefan law,
$\rho_r = aT^{4}$. As a consequence, one may check that the
product $Ta^{1 - \alpha/4}$ remains constant and, as such, this
implies that the new temperature law scales with redshift as
\begin{equation}
T = T_o (1 + z)^{1 - \alpha/4}.
\end{equation}

By inserting (\ref{energyR}) into (\ref{ECL}) it follows that
\begin{equation}
\rho_v = {\tilde{\rho}_{vo}} + \frac{\alpha\rho_{ro}}{4-\alpha}
a^{-4+\alpha},
\end{equation}
which should be compared with Eq. (\ref{decayv}) describing a
decaying vacuum energy density into cold dark matter. Note that
the ratio between the vacuum and radiation energy densities are:
\begin{equation}
\frac{\rho_v}{\rho_r} = \frac{\tilde{\rho}_{vo}}{\rho_{ro}} a^{4 -
\alpha} + \frac{\alpha}{4 - \alpha}.
\end{equation}
The first term is asymptotically vanishing  at early times whereas
the second one is smaller than unity. Therefore, a radiation
dominated stage is always guaranteed in this kind of scenarios.

{\bf Acknowledgments:} This work was supported by CNPq (Brazilian
Research Agency). The authors thank R. Silva, R. C. Santos and J.
F. Jesus for valuable discussions and Joan Sol\`a and Hrvoje Stefancic 
for pointing a error in the first version of this paper.

\end{document}